
%
\input phyzzx
\def\sp(#1){ \noalign{\vskip #1pt} }
\font\BIGr=cmr10 scaled \magstep2
\Pubnum={
INS-969 \cr
UT-633
}
\date={February 1993}
\titlepage
\title{\bf Transfer Matrix Formalism for Two-Dimensional
	 Quantum Gravity and Fractal Structures of Space-time }
\author{ H. KAWAI$^{\star}$, N. KAWAMOTO$^{\dagger}$,
T. MOGAMI$^{\ddagger}$  and Y. WATABIKI$^{\dagger}$ }
\vskip 16pt
\address{$\star$ Department of Physics, University of Tokyo,	     \break
              Bunkyo-ku, Tokyo 113, Japan                            \break
         $\dagger$ Institute for Nuclear Study, University of Tokyo, \break
              Tanashi, Tokyo 188, Japan                              \break
         $\ddagger$ Department of Physics, Kyoto University,         \break
              Kitashirakawa, Kyoto 606, Japan }
%
%
%
%
%
\abstract{ We develop a transfer matrix formalism for two-dimensional pure
 gravity.
By taking the continuum limit, we obtain a \lq\lq Hamiltonian formalism''
 in which the geodesic distance plays the role of time.
Applying this formalism, we obtain a universal
 function which describes the fractal structures of two dimensional
 quantum gravity in the continuum limit.
}
%
%
\endpage             
%
%
%

%
\REF\KPZDDK{
             V. G. Knizhnik, A. M. Polyakov and A. A. Zamolodchickov,
                Mod.\ Phys.\ Lett.\ {\bf A3} (1988) 819;
\nextline
             F. David,
                Mod.\ Phys.\ Lett.\ {\bf A3} (1988) 651;
\nextline
             J. Distler and H. Kawai,
                Nucl.\ Phys.\ {\bf B321} (1989) 509.
}
\REF\DT{
             F. David,
                Nucl.\ Phys.\ {\bf B257} (1985) 45;
\nextline
             V. A. Kazakov,
                Phys.\ Lett.\ {\bf 150B} (1985) 282;
\nextline
             D. V. Boulatov, V. A. Kazakov, I. K. Kostov and A. A. Migdal,
                Nucl.\ Phys.\ {\bf B275} (1986) 641;
\nextline
             J. Ambj\o rn, B. Durhuus and J. Fr\"ohlich,
                Nucl.\ Phys.\ {\bf B257} (1985) 433.
	     }
\REF\Matrix{
             E. Br\'ezin and V. A. Kazakov,
                Phys.\ Lett.\ {\bf 236B} (1990) 144;
\nextline
             M. Douglas and S. Shenker,
                Nucl.\ Phys.\ {\bf B335} (1990) 635;
\nextline
             D. J. Gross and A. A. Migdal,
                Phys.\ Rev.\ Lett.\ {\bf 64} (1990) 127.
            }
\REF\BIPZ{
             E. Br\'ezin, C. Itzykson, G. Parisi and J. B. Zuber,
                Commun.\ Math.\ Phys.\ {\bf 59} (1978) 35.
            }
\REF\KN{
             H. Kawai and M. Ninomiya,
                Nucl.\ Phys.\ {\bf B336} (1990) 115.
	     }
\REF\AM{
             M. E. Agishtein and A. A. Migdal,
                Int.\ J.\ Mod.\ Phys.\ {\bf C1} (1990) 165;
                Nucl.\ Phys.\ {\bf B350} 690 (1991).
	     }
\REF\KKSW{
             N. Kawamoto, V. A. Kazakov, Y. Saeki and Y. Watabiki,
                Phys.\ Rev.\ Lett.\ {\bf 68} (1992) 2113;
                Nucl.\ Phys.\ {\bf B}(Proc. Suppl.){\bf 26} (1992) 584.
             }
\REF\TY{
             N. Tsuda and T. Yukawa,
                private communications.
             }
\REF\KKMSW{
             H. Kawai, N. Kawamoto, T. Mogami, Y. Saeki and Y. Watabiki,
                in preparation.
             }
\def\Refmark(#1){$\,$[#1]}
\def\CMPrefmark(#1){ [#1]}

%
%
\topskip 30pt
Recent developments of two-dimensional gravity have provided us with an
 unambiguous definition of quantum gravity.
This is based on the equivalence of the continuum
 formulation\NPrefmark{\KPZDDK}
 and the dynamical triangulation.\NPrefmark{\DT}
In two dimensions, we thus have a regularized quantum gravity
 which has a definite continuum limit.
The remarkable success of the matrix models
\NPrefmark{\Matrix} further elucidated topological aspects of the theory.
 However we still lack a general formulation for describing
quantum fluctuations of space-time and for evaluating physical
 observables such as fractal dimensions.
In this paper we propose a new formulation which
is a kind of Hamiltonian
 formalism for quantum gravity.
We show that a geodesic distance defined on a dynamically triangulated
 surface can be regarded as the \lq\lq time'' variable for defining
the transfer matrix.
We then obtain a \lq\lq Hamiltonian'' in the continuum limit,
and analyze the fractal structures of the space-time.
\par
Let us consider a cylinder with an entrance loop($c$) and an exit loop($c'$).
 (See Fig. 1.)
We introduce the following quantity which is formally defined in the
 continuum framework:
$$ \eqalign{ \sp (2.0)
 N(L,L';D;A) \ = \ \int {Dg \over Vol(Diff)}
                & \delta(\int d^2x \sqrt g - A)
                  \delta(\int_{c} \sqrt{g_{\mu\nu}dx^\mu dx^\nu} - L)
\cr
                & \delta(\int_{c'} \sqrt{g_{\mu\nu}dx^\mu dx^\nu} - L')
                  \prod_{P\in c'} \delta(d(P,c) - D).
\cr
\sp(3.0)} \eqno(1) $$
Here in the path integral the total area of the surface
and the lengths of $c$ and $c'$
 are constrained to $A$, $L$ and $L'$, respectively.
We further impose a constraint that any point on $c'$ has a geodesic
 distance $D$ from the entrance loop $c$.
To be precise, the geodesic distance $d(P,c)$ is defined as the minimum
 distance between the point $P$ and a point on the loop $c$.
Note that the exit loop $c'$ is {\it one} of the loops which are
 composed of points having geodesic distance $D$ from the entrance loop $c$.
For later convenience we introduce a making point
 on the exit loop as is shown in Fig.1.
\par
An important property of $N(L,L';D;A)$ is the following composition law:
$$ \eqalign{ \sp(2.0)
 N(L,L';D;A) \ = \ \int_0^\infty dA' \int_0^\infty dL''
                    N(L,L'';D';A')N(L'',L';D-D';A-A').
\cr
\sp(3.0)} \eqno(2) $$
In other words, if we define $ N(L,L';D)$ by $ N(L,L';D) \, = \, \int_0^\infty
dA
N(L,L';D;A) e^{-tA}$, where $t$ is the cosmological constant,
it can be regarded as the matrix element $<L\mid e^{-D\hat H}\mid L'>$
 for a \lq\lq Hamiltonian'' $\hat H$.
In this sense we call $N(L,L';D)$ the proper time evolution kernel.
Roughly speaking, eq.(2) asserts that a cylinder with height $D$ can
 be decomposed into two cylinders with height $D'$ and $D-D'$.
More strictly, there could be several loops which have
 geodesic distance $D'$ from the entrance loop $c$.
{}From only one of them, however, the exit loop $c'$
has geodesic distance $D-D'$.
(See Fig.2)
\par
We now give a constructive definition of $N(L,L';D;A)$ in terms of
 dynamical triangulation.
The strategy to define its lattice counterpart
 $ \bar N(l,l';d;n)$ is the following.
We first introduce the notion of deforming a loop one step forward,
 whose precise definition is given in the next paragraph.
As we will see there, a loop can split into several loops after a one-step
 deformation.
In that case each loop is independently deformed in the next step.
By repeating this procedure $d$ times,
 we have a $d$-step deformation of a loop.
Then $ \bar N(l,l';d;n)$ is defined as the number of possible
 triangulations
 of a cylinder satisfying the following conditions which correspond to
 the four $\delta$-functions in eq.(1):
(i) The number of triangles is $n$.
(ii) The entrance loop ($c$) consists of $l$ links.
(iii) The exit loop ($c'$) consits of $l'$ links.
(iv) The exit loop is one of the connected components obtained after
 a $d$ step deformation of the entrance loop.
\par
In order to give a precise definition of the deformation, let us consider
 a loop $c$ on a triangulated surface.
To deform $c$ one step forward means to remove triangles attached to $c$
 in the forward direction and two-fold links on $c$.
%
%
 A typical example is illustrated in Fig.3, where $c$ is represented
by a solid line, and the forward direction is assumed to be
inward. In this example, $c$ has three two-fold links
on it as indicated by $\alpha , \beta$ and $\gamma$,
but we still regard it as a single loop.
 After a one-step deformation, this $c$ will split into three loops
which are indicated by the dotted lines.
 The notion of two-fold links might seem an idle complexity.
 As we will see, however, it plays an important role in the explicit
evaluation of the transfer matrix.
Note that the deformation of a loop considered here is closely related
 to the geodesic distance on the dual lattice.
Actually, if a loop $c'$ is obtained from loop $c$ after a $d$-step
 deformation, any point on $c'$ has geodesic distance $d$ from $c$.
This is why we impose the condition (iv) in order to express the last
 $\delta$-function in eq.(1).
\par
Since $\bar N(l,l';d;n)$ is the lattice counterpart of $N(L,L';D;A)$, it
 satisfies the same composition law as eq.(2).
Therefore, if we define the $d$-step evolution kernel $\bar N(l,l';d)$ by
$ \bar N(l,l';d) \ = \ \sum^\infty_{n=0} \bar N(l,l';d;n)K^n $
and regard it as the $l$, $l'$ element of a matrix $\hat N(d)$,
it can be decomposed into a product of single step evolution kernels:
$$ \eqalign{ \sp(2.0)
 \bar N(l,l';d) \ \equiv \ (\hat N(d))_{l,l'} \ = \ (\hat N(1)^d)_{l,l'},
\cr
\sp(3.0)} \eqno(3) $$
which means that $\hat N(1)$ plays the role of a transfer matrix.
\par
Next we show that the transfer matrix $\hat N(1)$ can be evaluated by
 a combinatorial analysis.
To proceed, we first need to evaluate the disk amplitude
$$ \eqalign{ \sp(2.0)
 \tilde F(y;K) \ = \ \sum_{l,n=0}^\infty  y^l K^n F(l;n),
\cr
\sp(3.0)} \eqno(4) $$
where $F(l;n)$ is the number of possible triangulations of a disk which
 has a boundary of length $l$ and consists of $n$ triangles.
The disk amplitude $\tilde F(y;K)$ can be evaluated by the large-$N$
 $\phi^3$ matrix model\NPrefmark{\BIPZ} as
$$ \eqalign{ \sp(2.0)
 \tilde F(y;K) \ = \ {1\over y}
	 \Big\{
	 {1\over 2} ( {1\over y} -{K\over y^2} )
	+ {K\over 2}({1\over y}-c)\sqrt{({1\over y}-a)({1\over y}-b)}
	 \Big\},
\cr
\sp(3.0)} \eqno(5) $$
where $a$, $b$ and $c$ are functions of $K$ determined by
$$ \eqalign{ \sp(2.0)
      a &= {1\over K}-c + \sqrt{2c({1\over K}-c)} ,
\cr
      b &= {1\over K}-c - \sqrt{2c({1\over K}-c)},
\cr
      {1\over K} &= c({1\over K}-c)(c-{1\over 2K})     .
\cr
	& (~b<0<a<c~)
\cr
\sp(3.0)} \eqno(6) $$
As is easily seen from (5) and (6), $\tilde F(y;K)$ is analytic in both
 $K$ and $y$ with finite convergence radii.
At the critical point where $\tilde F(y;K)$ becomes singular as a
 function of $K$, the values of $a$ and $c$ coincide and the convergence
 radius with respect to $y$ is equal to $1/a$.
These critical values can be easily obtained from eq.(6):
$K_c^2 \,= \, 1/12\sqrt3, \ y_c \, = \, (3^{1/4}-3^{-1/4})/2, \
a_c \, = \, c_c \, = \, 1/y_c, \ b_c \, = \, (1-\sqrt3)/{2K_c}$.
\par
Near this critical point we take the continuum limit by parametrizing
 $K$ and $y$ as
$$ \eqalign{ \sp(2.0)
 K \, &= \, K_c e^{-\epsilon^2t},~~~~~~~~~~~~~~~~~~~~~~~
 y \, = \, y_c e^{-\epsilon\zeta} .
\cr
\sp(3.0)} \eqno(7) $$
Then $a$, $b$ and $c$ are calculated from eq.(6) up to order $\epsilon$ as
$$ \eqalign{ \sp(2.0)
 a \, &= \, a_c - {4\over 3^{1/4}} \epsilon\sqrt t,~~~
 b \, = \, b_c,~~~
 c \, = \, c_c + {2\over 3^{1/4}} \epsilon\sqrt t.
\cr
\sp(3.0)} \eqno(8) $$
We point out that the parametrization of eq.(7) is a natural one
 in the following sense.
If we replace $K^n$ and $y^l$ in eq.(4) with
$K^n \, = \, K_c^n e^{-n\epsilon^2t}$ and
$y^l \, = \, y_c^l e^{-l\epsilon\zeta}$, respectively,
it is clear that the continuum limit $\epsilon \rightarrow 0$ corresponds
 to taking the $n \rightarrow \infty$
 and $l \rightarrow \infty$ limit with
 the physical area $A \, = \, n \epsilon^2$ and the physical length
 $L=l\epsilon$ kept finite.
Thus $t$ and $\zeta$ are the conjugate variables to the area and the
 boundary length, respectively.
\par
Substituting eqs.(7) into eq.(5), we obtain the following disk amplitude
 near the continuum limit:
$$ \eqalign{ \sp(2.0)
\tilde F(\zeta;t) \, = \,{1\, + \, \sqrt3 \over 2}(1 \,
 - \, \sqrt3 \epsilon\zeta)
 \, + \, {(1 \, + \,\sqrt3)^{5/2}\over 4}f(\zeta,\tau)\epsilon^{3\over2}
 \, + \, O(\epsilon^2),
\cr
\sp(3.0)} \eqno(9) $$
where
$$ \eqalign{ \sp(2.0)
f(\zeta,\tau) \ = \ (2\zeta - \sqrt \tau)\sqrt{\zeta +\sqrt \tau},
{}~~~~~~~
\sqrt \tau \ = \ {4\over 3 \, + \, \sqrt3}\sqrt t.
\cr
\sp(3.0)} \eqno(10) $$
Here $f(\zeta,\tau)$ is a universal function in the sense that it does
 not depend on the details of the triangulation prescription.
\par
We now use combinatorics to evaluate the generating function of the
 matrix element of the transfer matrix $\hat N(1)$:
$$ \eqalign{ \sp(2.0)
 \tilde N(y,y';K) \ \equiv \ \sum_{l,l'}^\infty y^{l}y'^{l'}
   \big(\hat N(1)\big)_{l,l'}
  \ \equiv \ \sum_{l,l,n=0}^\infty y^{l}y'^{l'} K^n \bar N(l,l';1;n).
\cr
\sp(3.0)} \eqno(11) $$
What we are going to do is to sum up all possible
triangulations of a cylinder such
 that the exit loop $c'$ is one of the connected components of the
 one-step deformation of the entrance loop $c$.
In order to define
a natural matrix multiplication of the transfer matrix,
 we introduce a marking point on $c'$ but not on $c$.(See Fig.4.)
Let us first consider the triangle which is attached to the marked link.
There are only three possible types for this marked triangle; 1), 2),
 and 3)  in Fig.5a, where $F$ in 3) denotes an arbitrary triangulation
 of a disk which is connected to the marked triangle through a vertex.
After a one-step deformation, $c$ proceeds to the marked link on $c'$ and
 the disk part will be disconnected.
As we can see in eq.(11), the power of $y$, $y'$, and $K$ counts the
 number of links on $c$, links on $c'$, and triangles respectively.
Then the contribution to $\tilde N(y,y';K)$ from the marked triangles
 is easily evaluated as follows: 1) $yy'^2K$, 2) $yy'^2K$,
 3) $y^2 y'K \tilde F(y;K)$.
\par
We then examine the structure of the rest of the triangulation by
 going along the exit loop $c'$.
There are four different types of basic structures 1),2),3) and 4)
 as depicted in Fig.5b.
Their weights are easily evaluated as 1) $yy'^2K$,
 2) $y^2y'K\tilde F(y;K)$, 3)
 $y^2\tilde F(y;K)$, and 4) $yK(\tilde F(y;K)-1)$.
Starting with the marked triangle, we can attach any of these four
 structures one by one repeatedly, and then come back to the original
 marked triangle.
Therefore the contribution of these four structures to $\tilde N(y,y';K)$
 can be expressed by a geometric series.
We thus obtain the following form for the generating function of the
 transfer matrix elements:
$$ \eqalign{ \sp(2.0)
 \tilde N(y,y';K) \ &= \ \{ 2yy'^2K  +  y^2 y' K \tilde F(y;K) \}
 \cr
             & \ \ \times \sum^\infty_{n=0} \{yy'^2K  +  y^2y'K\tilde F(y;K)
	     +  y^2\tilde F(y;K)  +  yK(\tilde F(y;K)-1) \}^n
 \cr
             &= \ { 2yy'^2K  +  y^2 y' K \tilde F(y;K) \over
             1  -  yy'^2K  -  y^2y'K\tilde F(y;K)  -  y^2\tilde F(y;K)
		  -  yK(\tilde F(y;K)-1)}.
 \cr
\sp(3.0)} \eqno(12) $$
\par
We next consider the continuum limit of this transfer matrix.
In order to see how the continuum limit should be taken, let us express
 the composition law in terms of the generating functions:
$$ \eqalign{ \sp(2.0)
  {1 \over 2\pi i} \oint {dz \over z}
              \tilde N(y,z;d_1;K) \tilde N({1\over z},y';d_2;K)
              \ = \ \tilde N(y,y';d_1+d_2;K),
\cr
\sp(3.0)} \eqno(13) $$
where
$\tilde N(y,y';d;K) = \sum_{l=0}^\infty \sum_{l'=0}^\infty \sum_{n=0}^\infty
y^{l}y'^{l'} K^n \bar N(l,l';d;n)$.
{}From eq.(13) it is clear that the continuum limit for $y$ and $y'$
 should be taken around different values which are inverse to each other.
Furthermore, since the structure of the entrance loop is similar to
 the boundary loop of the disk amplitude, it is natural to expect that
 the continuum limit can be taken by setting
$$ \eqalign{ \sp(2.0)
               y \ = \ y_ce^{-\epsilon \zeta},~~~~~
               y' \ = \ y_c^{-1}e^{-\epsilon \zeta'}, ~~~~~
               K \ = \ K_c e^{-\epsilon^2 t}.
\cr
\sp(3.0)} \eqno(14) $$
If this is true, the composition law (13) can be re-expressed as
$$ \eqalign{ \sp(2.0)
       {\epsilon \over 2\pi i}
       \int_{-i\infty}^{i\infty} d\xi
       \tilde N(\zeta,\xi;d_1;t) \tilde N(-\xi,\zeta';d_2;t)
  \ = \  \tilde N(\zeta,\zeta';d_1+d_2;t),
\cr
\sp(3.0)} \eqno(15) $$
where $ \tilde N(\zeta,\zeta';d;t)$ stands for $\tilde N(y,y';d;K)$ for
 the values of $y$, $y'$ and $K$ given by (14).
\par
The validity of this continuum limit is explicitly checked
 by substituting (14) into eq.(12).
We obtain
$$ \eqalign{ \sp(2.0)
  \tilde N(y,y';K) \ \equiv \ \tilde N(\zeta, \zeta';t) \ = \
  {1 \over \epsilon} {1 \over \zeta' \, + \, \zeta \, - \,
  \, \alpha \epsilon^{1\over 2} f(\zeta, \tau) } \ + \ O(\epsilon^0),
\cr
\sp(3.0)} \eqno(16) $$
where $\alpha \, = \, \sqrt{2/(9\sqrt3 - 1)}$ and $f(\zeta, \tau)$ is
 given by eq.(10).
In this eqation we have
the following miraculous cancellations, which convince us that
 the continuum limit we consider here is in fact the right one.
First of all $O(\epsilon^0)$ terms in the denominator of
 $\tilde N(y,y';K)$
 cancel out. Secondly, the coefficients of $\zeta$ and $\zeta'$ are 1.
Thirdly,
 the residue of $\tilde N(y,y';K)$ with respect to the $\zeta + \zeta'$
 term is $1/\epsilon$, which, as we see below, exactly cancels
 the factor $\epsilon$ in (15).
These are highly nontrivial results.
\par
By substituting (16) into (15) for $d_1=1$ and $d_2=d$, we obtain
$$ \eqalign{ \sp(2.0)
\tilde N(\zeta,\zeta';d+1;t) \
    &= \ {1 \over 2\pi i} \int_{-i\infty}^{i\infty} d\xi
    {1 \over \zeta \, + \, \xi \, - \,
    \, \alpha \epsilon^{1\over 2} f(\zeta, \tau) }
    \tilde N(-\xi,\zeta';d;t) \ ,
\cr
    &=\  \tilde N(\zeta-
     \alpha \epsilon^{1\over 2} f(\zeta, \tau),\zeta';d;t),
\cr
\sp(3.0)} \eqno(17) $$
which leads to
$$ \eqalign{ \sp(2.0)
 \tilde N(\zeta,\zeta';d+1;t) \ - \ \tilde N(\zeta,\zeta';d;t) \ = \ -
                       \alpha \sqrt\epsilon f(\zeta, \tau)
		       {\partial \over \partial \zeta}
                       \tilde N(\zeta,\zeta';d;t).
\cr
\sp(3.0)} \eqno(18) $$
We then take the continuum limit of eq.(18) by introducing
$D \equiv \alpha \sqrt\epsilon d $ and taking
$\epsilon \rightarrow 0$.
We thus obtain the following continuum differential equation for
$\psi(\zeta;D) =  \tilde N(\zeta,\zeta';D;t)$:
$$ \eqalign{ \sp(2.0)
 {\partial \over \partial D} \psi(\zeta,D) \ = \
         - f(\zeta, \tau) {\partial \over \partial \zeta} \psi(\zeta,D).
\cr
\sp(3.0)} \eqno(19) $$
In other words, the proper time evolution kernel can be identified with
 the following matrix element:
$\tilde N(\zeta,\zeta';D;t) = <\zeta\mid e^{-D H} \mid \zeta'>$,
where $ H$ is a \lq\lq Hamiltonian'' given by
$$ \eqalign{ \sp(2.0)
        \ H \ = \ f(\zeta,\tau) {\partial \over \partial \zeta}.
\cr
\sp(3.0)} \eqno(20) $$
It should be noted that the miraculous cancelations that occurred
 in eq.(16) are crucial in the derivation of the continuum differential
 equation (19).
\par
The initial value problem for the differential equation (19) can be easily
 solved by the use of the characteristic curve method as
$\psi(\zeta;D) = \psi(\zeta';0)$, where $\zeta'$ is defined by
$ \int^\zeta_{\zeta'} d\zeta'' / f(\zeta'',\tau) \ = \ D $
 as a function of $\zeta$ and $D$.
We can now evaluate the proper time evolution kernel $N(L,L';D)$ which is
 given by the inverse Laplace transformation of
 $\tilde N(\zeta,\zeta';D;\tau)$ for $\zeta$ and $\zeta'$.
\foot{
$N(L,L';D)$ is related to its discrete version $\bar N(l,l';d)$ by
$ N(L,L';D) = \epsilon^{-1} {y_c}^{l-l'} \bar N(l,l';d)$.
}
By carrying out an inverse Laplace transformation for the solution
 of (19) with the initial condition $\psi(\zeta;0) = e^{-L'\zeta}$, we obtain
$$ \eqalign{ \sp(2.0)
    N(L,L';D) \ = \ {1 \over 2\pi i} \int_{-i\infty}^{i\infty} d\zeta
	      e^{L\zeta} e^{-L'\zeta'}.
\cr
\sp(3.0)} \eqno(21) $$
Note that eq.(21) indeed reproduces the corresponding initial condition
$N(L,L';0) = \delta(L-L')$,
because we have $\zeta' = \zeta$ for $D=0$.
\par
For later use we give $N(L,L';D)$ for small values of $L$ and $\tau$:
$$ \eqalign{ \sp(2.0)
    N(L,L';D) \ =& \ {L' \over \sqrt{\pi L}} \Big\{ {2 \over D^3}
                      -\tau{3\over10}(D + {L' \over D}) + \tau^{3\over2}
		      {1\over7}(D^3 + {1\over2}L'D) + O(\tau^2) \Big\}
		      e^{-L'/D^2} \
\cr
                   &+ \ O(L^0).
\cr
\sp(3.0)} \eqno(22) $$
We also calculate the inverse Laplace transformation
 of the disk amplitude as a double series expansion
 with respect to $\epsilon$ and $t$:
$$ \eqalign{ \sp(2.0)
    F(L) \ &= \ {1 \over 2\pi i} \int_{-i\infty}^{i\infty} d\zeta \
	        e^{L\zeta} \ \tilde F(\zeta;t) \epsilon^{-3/2}
\cr
             &= \ {1+\sqrt3 \over 2}\delta (L-\sqrt 3\epsilon)\epsilon^{-3/2}
\cr
             &~~~~+{3(1+\sqrt3)^{5/2}\over 8\sqrt\pi}\Big( L^{-5/2} -
                {\tau\over 2} L^{-1/2} + {\tau^{3/2} \over 3} L^{1/2} \Big)
                 + O(\tau^2),
%
\cr
\sp(3.0)} \eqno(23) $$
where $\tilde F(\zeta;t)$ is given by eq.(9).
\par
We now apply the formalism developed here to the analysis
 of the fractal structures of the space-time.\NPrefmark{\KN,\AM,\KKSW}
We consider a large enough space-time with spherical topology
 for two-dimensional pure gravity.
We take an arbitrary point $P$ and consider the set of points $S(P;D)$
 whose geodesic distances from $P$ are less than or equal to $D$.
The boundary of $S(P;D)$ usually consists of many loops with various lengths.
Let $\rho(L;D)dL$ be the number of loops belonging to the boundary of
 $S(P;D)$ whose lengths lie between $L$ and $L+dL$.
As we explain below, the quantity $\rho(L;D)$ can be evaluated as
$$ \eqalign{ \sp(2.0)
     \rho(L;D) \ &= \ \lim_{L_0\rightarrow 0, \tau \rightarrow 0}
                {{\partial^2 \over \partial \tau^2}
                N(L_0,L;D){1\over L} F(L) \over
                {\partial^2 \over \partial \tau^2} {1\over L_0} F(L_0)},
\cr
              &= \, {1\over D^2}G(x)
	\, + \, {4 \, \epsilon^{-3/2} \over 7 (1+\sqrt{3})^{3/2}} D^3
	(1-{x \over 2})
	\delta(L-\sqrt{3} \epsilon),
\cr
\sp(3.0)} \eqno(24) $$
where
$$ \eqalign{ \sp(2.0)
        G(x) \ \equiv \ {3\over 7 \sqrt{\pi}}(x^{-5/2} + {1\over2} x^{-3/2}
 + {14\over3} x^{1/2})
                         e^{-x},
\cr
\sp(3.0)} \eqno(25) $$
with $x = L/D^2$ as a scaling parameter.
 In eq.(24), the limit $L_0 \rightarrow 0$ corresponds to shrinking
the entrance loop to a point $P$, and the limit $\tau \rightarrow 0$
is taken to have the thermodynamic ilmit.
The second derivatives with respect to $\tau$ in eq.(24) are introduced
 to avoid small area dominance in the $\tau \rightarrow 0$ limit.
Any higher derivative works for this purpose without changing the
 final answer.
The numerator on the RHS corresponds to an operation of gluing a disk
 with a boundary of length $L$ to the exit loop of the cylinder.
The factors $1/L_0$ and $1/L$ in front of $F$'s are introduced to
 convert marked boundaries to unmarked ones.
As we have already explained, there are several loops which have
 geodesic distance $D$ from the entrance loop.
Recognizing that the exit loop is one of these loops, we easily
 realize that the ratio in eq.(24) counts the number of loops which have
 the boundary length $L$.
\par
One of the surprising properties of the function $\rho(L;D)$ is that
 it is essentially a universal scaling function of the scaling
 parameter $x = L/D^2$.
For small values of $L$, however, $\rho(L;D)$ includes non-universal
 parts which have negative power dependence on the lattice
 constant $\epsilon$.
In order to examine the scaling property, it is convenient to
 introduce the following quantities:
$$ \eqalign{ \sp(2.0)
     <L^n> \ = \ \int_0^\infty dL ~ L^n \rho(L;D).
\cr
\sp(3.0)} \eqno(26) $$
{}From eq.(24) it is easy to show that
$$ \eqalign{ \sp(2.0)
     <L^0> \ &\propto \ const \times D^3 \epsilon^{-3/2}
     	+const \times D \epsilon^{-1/2}
                      + const \times D^0,
\cr
     <L^1> \ &\propto \ const \times D^3 \epsilon^{-1/2}
                      + const \times D^2,
\cr
     <L^n> \ &\propto \ const \times D^{2n} ~~~~~(n\geq2).
\cr
\sp(3.0)} \eqno(27) $$
\par
Since $D^2 \rho(L;D)$ is essentially
 the universal function $G(x)$ of the
 dimensionless parameter $x$, it is an ideal quantity to measure in
 computer simulations.
In fact a numerical study for pure gravity shows a clear universal
 scaling behavior according to $G(x)$.\NPrefmark{\TY}
As we can see in eq.(27), $<L^0>$ and $<L^1>$ include $\epsilon$
 dependent non-universal part as a dominant contribution while
 $<L^n>$ for $n \geq 2$ includes only universal part and thus is
 expected to show clear fractal behaviors.
The controversial numerical results for the fractal dimension by
 Agishtein and Migdal\NPrefmark{\AM} can be understood by the
 non-universal behavior of the corresponding quantities.
\par
Although we have only treated pure gravity ($c=0$) in this paper,
 it is important to investigate quantum gravity coupled to matter fields.
In the limit $c \rightarrow -\infty$ it is known that the two-dimensional
 gravity system approaches to the classical limit.
We thus expect that the wildly branching space-time surface approaches
 to a smooth surface.
On the smooth surface there is only a single boundary
 at a geodesic distance $D$ from a given point.
We then have
$\rho(L;D) = \delta(L-2\pi D) = 1/D\delta (L/D - 2\pi)$,
which again suggests a scaling function with respect to a scaling
 parameter $x = L/D$.
It is thus very natural to expect that the gravity, for example,
 with $c=-2$ matter shows the similar scaling behavior.
In fact our recent numerical investigation shows a clear scaling
 behavior of the function $\rho(L;D)$ which has
 a similar scaling parameter
 $x=L/D^{1.9\pm 0.2}$ but has a slightly different power behavior from
 (25).\NPrefmark{\KKMSW}
It would be very interesting if we can find an analytic formulation
for deriving $\rho(L;D)$ in a model with $c \neq 0$.
\par
In this paper we have given an explicit form of \lq\lq Hamiltonian''
 by eq.(20) which could be viewed as a time evolution generator of
 closed string, where time is identified with the geodesic distance.
This formulation may provide a new formulation of closed string field theory.

%
%
%
%
\vskip 1cm    
%
\ack
We thank N. Tsuda and T. Yukawa for informing us of the results of their
 numerical simulations prior to publication.
We also thank Y. Okamoto for valuable comments.

%
%
%
\endpage
%
%
%
%
%
%
%
\endpage
%
%
\def\Figures{
\centerline{\BIGr FIGURE CAPTIONS}
\vskip 8pt
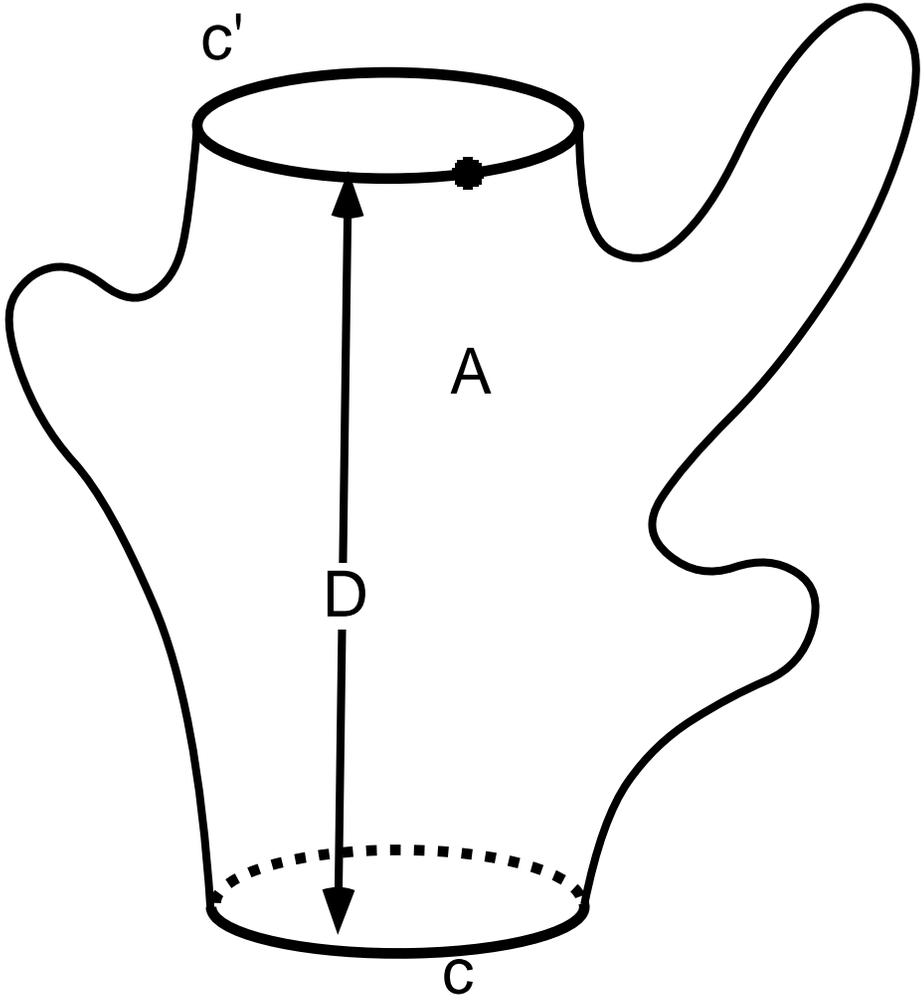
\item{\rm Fig.~1}{ Cylinder with entrance loop $c$ and exit loop $c'$.
}
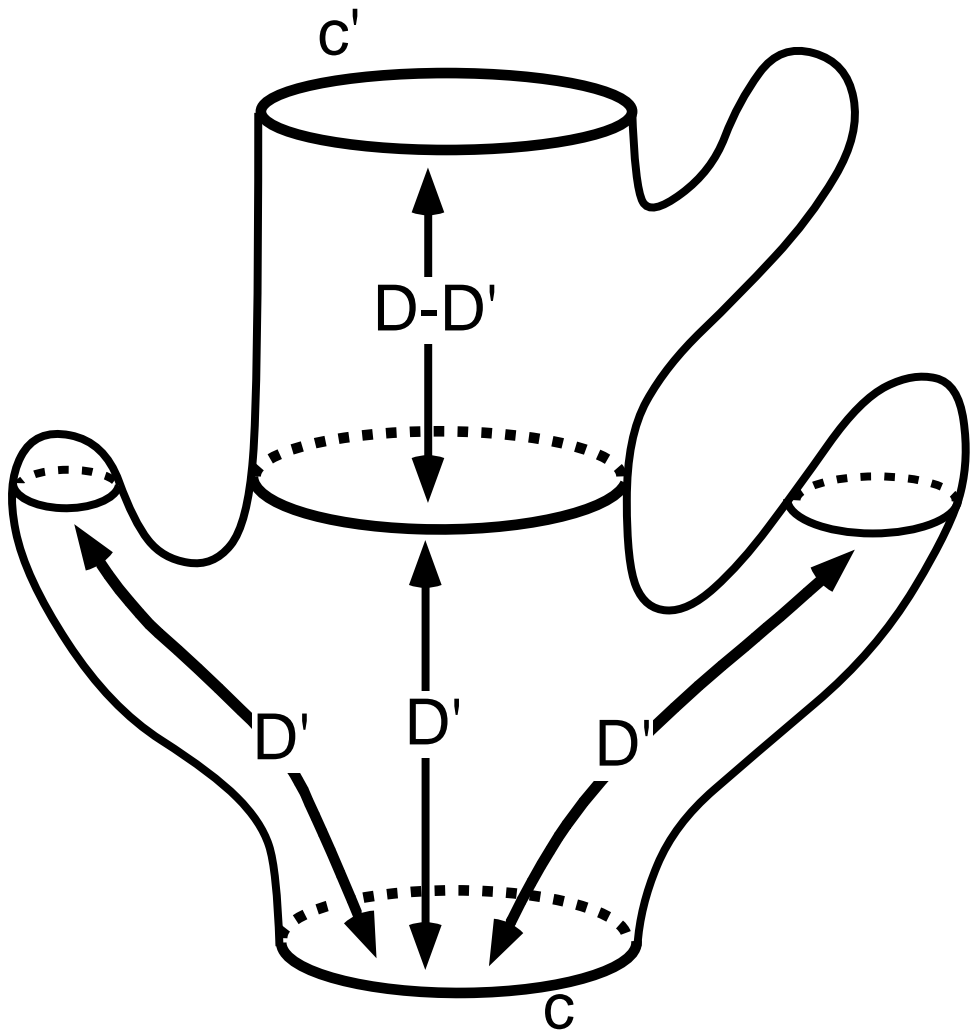
\item{\rm Fig.~2}{ Decomposition of a cylinder with height $D$ into two
 cylinders with height $D'$ and $D-D'$.
}
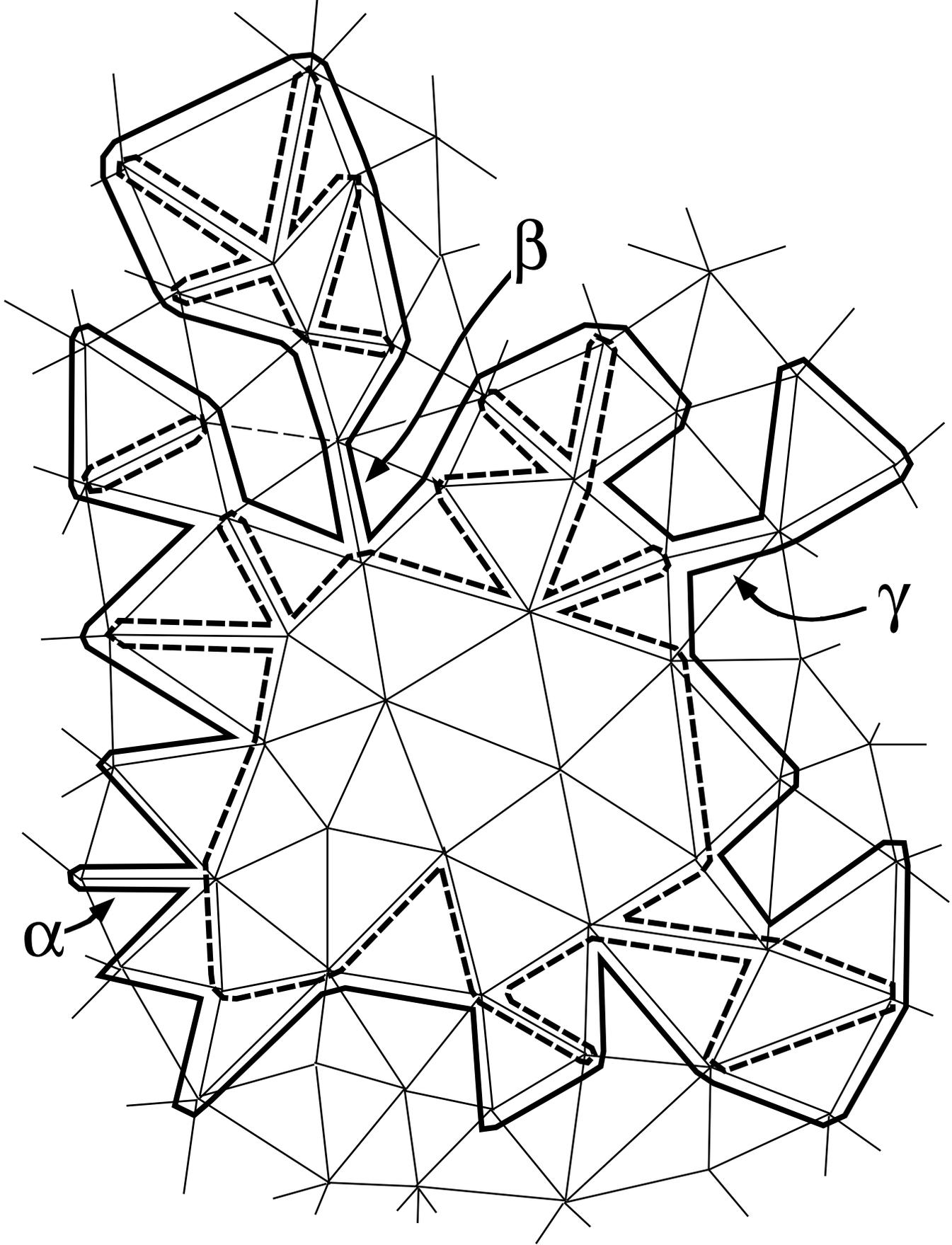
\item{\rm Fig.~3}{ A Typical example of the deformation of a loop.
 The dotted loops are obtained after
 a one-step deformation of the solid loop.
}
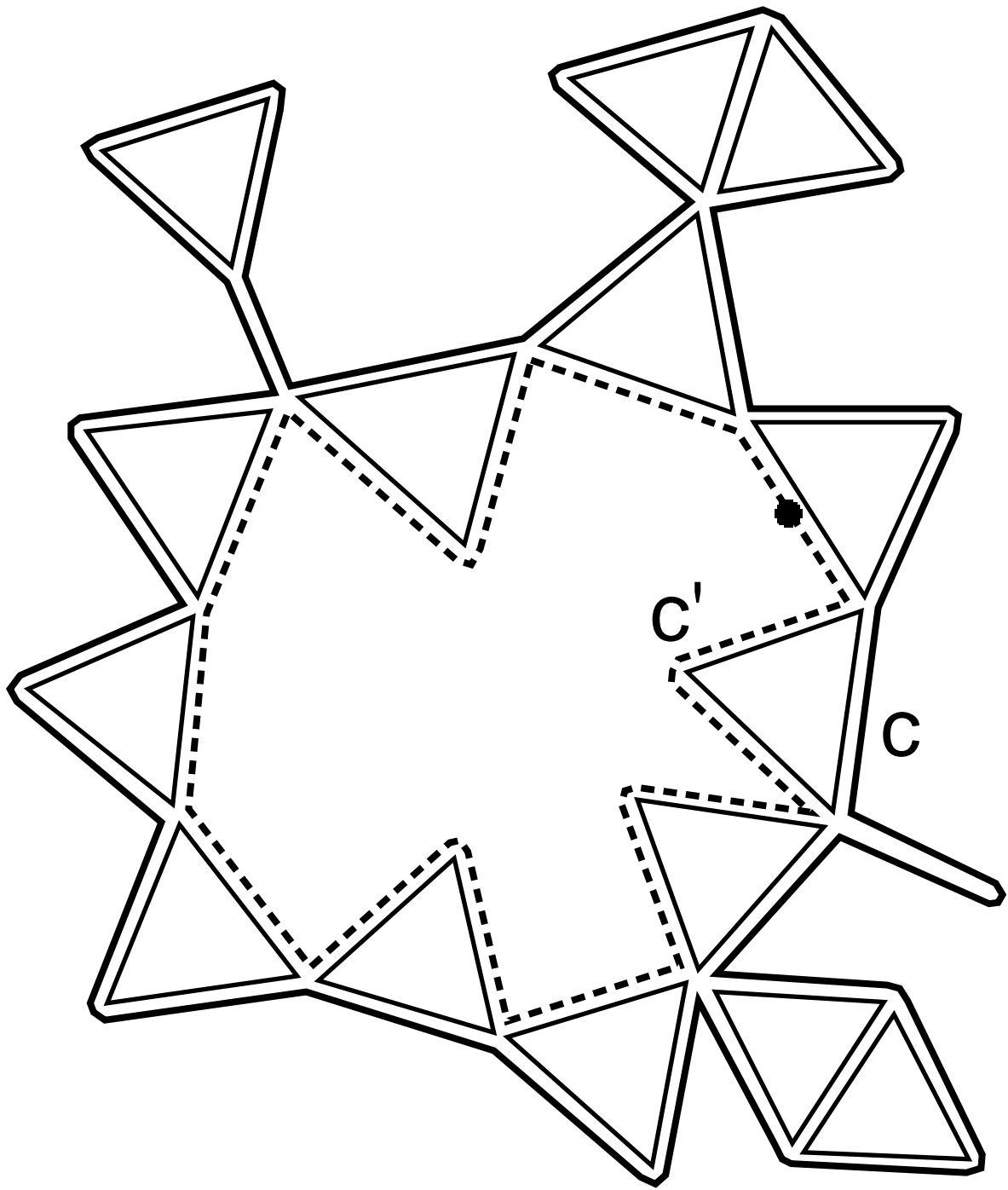
\item{\rm Fig.~4}{
 A triangulation contributing to $\tilde N(y,y';K)$.
The dotted loop $(c')$ is
one of the components of the one-step deformation of the solid line $(c)$.
}
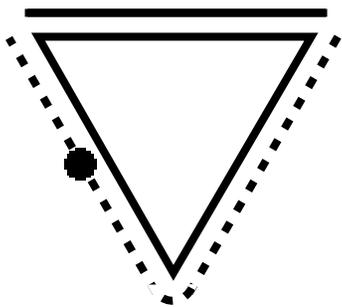
\item{\rm Fig.~5}{
a) Three types for the triangle attached to the marked link.
b)Four basic structures which may appear along $c$.
The solid and dotted lines stand for a part of the entrance($c$) and
 the exit($c'$) loops, respectively.
}
}
%
%
%
%
%
%
%
%
%
\refout
\vfill\eject
\Figures
%
%
%
\vfill
%
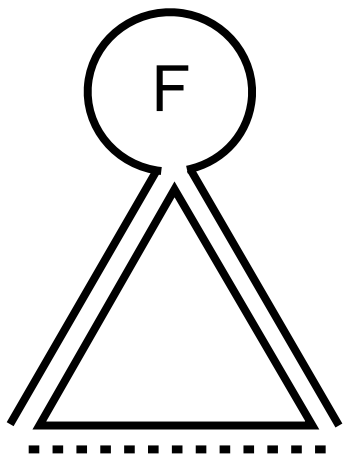
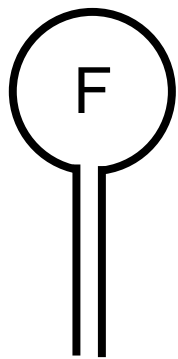
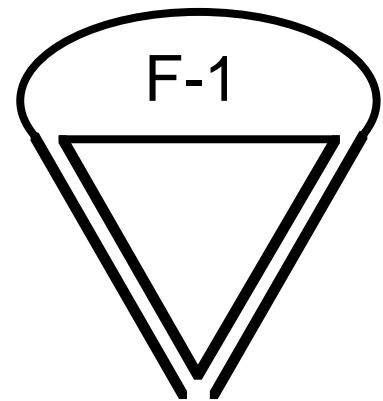
\bye